\documentstyle[prb,aps]{revtex}
\begin{document}
%\draft
\title{Coulomb Charging at  Large Conduction}
\author{Xiaohui Wang  and Hermann Grabert}
\address{Fakult{\"a}t f{\"u}r Physik, Universit{\"a}t Freiburg,
Hermann-Herder-Stra{\ss}e  3, 79104  Freiburg,  Germany}
\date{\today}
\maketitle
\begin{abstract}
We discuss the suppression of Coulomb charging effects on
a small metallic island coupled to an electrode by a tunnel
junction.  At high temperatures the quantum corrections
to the classical charging energy $E_c=e^2/2C$, where $C$
is the island capacitance, are evaluated. At low temperatures
the large quantum fluctuations of the island charge cause
a strong reduction of the effective $E_c$
which is determined explicitly in the limit
of a large tunneling conductance.
\end{abstract}


%\narrowtext
In recent years Coulomb blockade effects in metallic nano\-structures
have been
studied extensively both theoretically and experimentally\cite{gra92}.
These effects arise in systems with small metallic
islands separated by tunnel barriers.
For the case of weak tunneling addressed mostly so far\cite{ing92}
the transfer of charges can be described in terms of a tunneling
Hamiltonian treated perturbatively. On the other hand, for
strong tunneling one expects a washout of charging effects
by large quantum fluctuations of the charge.  For a quantum dot
coupled by a point contact to a lead electrode, the dependence
of charging effects on the transmission coefficient was
studied in recent work\cite{mat95}.  However, in this case
there are only few transport channels.

The behavior of metallic tunnel junctions with many channels
can conveniently be described by a path integral over
fluctuations of the phase conjugate to the island charge\cite{ben83}.
Based on this approach the suppression of Coulomb effects
was studied earlier by several authors\cite{gui86,pan91,zw91,fal95}.
Most of
these papers have considered single junctions while it become
clear later\cite{dev90} that in this case charging effects are
usually suppressed by the coupling to electromagnetic modes.
On the other hand, charge quantization on small metallic
islands is predominantly affected by fluctuations of charge
carriers across adjacent tunnel junctions
\cite{av92,mat91,gog94,gra94,sch94}.

The simplest device displaying these effects is the so--called
single electron box (SEB) which is formed by a metallic island
between by a tunnel junction and a gate capacitor (cf.\ Fig.~1).
The external
electrodes are biased by a voltage source\cite{laf93}.  Recently,
Falci {\it et al.}\cite{fal95} have examined the low temperature
behavior of the SEB by means of renormalization group techniques.
While powerful in predicting the main parameter dependence of
the effective charging energy in the low temperature range,
this method does not allow for a determination of the precise
relation between low temperature parameters and quantities
measurable in the classical regime. On the other hand, experiments
cover both the range of classical and quantum mechanical charge
fluctuations and one would like to examine the low temperature
behavior of devices in terms of parameters measured in the
classical regime.

In this article we present a new method to evaluate the
path integral in phase representation. At low temperatures
the path integral is dominated by ``sluggon'' trajectories
describing  phase creep by $2\pi$ across the junction. The
contribution of these trajectories can be evaluated explicitly
in the limit of low temperatures and large tunneling conductance.
As a main result we find that the dominant effect of charge
fluctuations can be absorbed in an effective charging energy.

We start from the partition function of the SEB at inverse
temperature $\beta=1/k_BT$ which may be written as
\[
Z=\int D [\varphi] e^{-S_{\rm Box}[\varphi]},
\]
where the  integral is over all paths of the phase
$\varphi(\tau)$ in the interval $-\beta/2 \leq \tau \leq \beta/2$
with $\varphi(\beta/2)=\varphi(-\beta/2)$ modulus  $2\pi$. The action
$S_{\rm Box}[\varphi]=S_c[\varphi]+S_t[\varphi]$
contains two parts describing charging of the island and
tunneling across the junction.  The effect of the Coulomb energy is
contained in
\[
S_c[\varphi]=\int_{-\beta/2}^{\beta/2} d\tau \left[ \frac
{1}{4E_c}(\dot{\varphi}+2in_{\rm ex}E_c)^2+E_cn_{\rm ex}^2 \right]
\]
where $E_c=e^2/2C$ is the single electron charging energy,
in which the island capacitance $C$ is the sum of the
capacitance $C_t$ of the tunnel junction and the
gate capacitance $C_g$.
$n_{\rm ex}=C_gU/e$ is a dimensionless measure of the
voltage $U$ applied via the gate capacitance (cf.\ Fig.~1).
The second part of the action\cite{ben83}
\[
S_t[\varphi]=2\int_{-\beta/2}^{\beta/2}
d\tau \int_{-\beta/2}^{\beta/2} d\tau' \alpha(\tau-\tau')
\sin^2 \left[ \frac{\varphi(\tau)-\varphi(\tau')}{2} \right]
\]
describes tunneling.
The  Fourier transform of $\alpha(\tau)$ reads
$\alpha_l=-\alpha_t |\omega_l|/4\pi$
for $|\omega_l| \ll D$
where the $\omega_l=2\pi l/\beta$ are
Matsubara frequencies, $D$ is
the electronic bandwidth, and  $\alpha_t=R_K/R_t$
is the dimensionless conductance of the junction,
where $R_t$  is the tunneling resistance and $R_K=h/e^2$
the von-Klitzing resistance.

Since the action is a periodic function of
$\varphi$ with period $2\pi$,  the partition function
may be written as a sum over  ``winding numbers'',
\[
Z=\sum_{k=-\infty}^{\infty} e^{2\pi i kn_{\rm ex}} Z_k .
\]
Here
\[
Z_k=\int_{\varphi(\beta/2)=\varphi(-\beta/2)+2\pi k}
D [\varphi] e^{-S[\varphi]}
\]
where the action
$S[\varphi]=S_0[\varphi]+S_t[\varphi]$,
with $S_0[\varphi]=\int_{-\beta/2}^{\beta/2}
d \tau {\dot{\varphi}}^2/4E_c $.

Due to the nonlinear, nonlocal interactions
described by $S_t[\varphi]$, the partition
function cannot be evaluated exactly.
Here we shall employ  semiclassical  methods.
The classical paths (from $\delta S[\varphi]=0$) satisfy
\begin{equation}
\ddot{\varphi}-4E_c\int_{-\beta/2}^{\beta/2}d\tau'
\alpha (\tau-\tau')
\sin[\varphi(\tau)-\varphi(\tau')] =0,  \label{clpath}
\end{equation}
and the boundary condition for winding number $k$ reads
$\varphi(\beta/2)=\varphi(-\beta/2)+2\pi k$.
A trivial solution of  this equation is\cite{zw91,fal95}
\[
\varphi_{\rm cl}^{(k)}(\tau)=\varphi(-\beta/2)+\omega_k
(\tau+\beta/2)
\]
with the classical action
\[
S_{\rm cl}^{(k)}=\frac{\pi^2 k^2}{\beta E_c}+
|k|\frac{\alpha_t}{2} .
\]
For an arbitrary  path $\varphi(\tau)=
\varphi_{\rm cl}^{(k)}(\tau)+\theta(\tau)$ of winding number $k$
fluctuating about the classical solution,
the second order variational action reads
\[
\delta^2 S^{(k)}= \int_{-\beta/2}^{\beta/2} d\tau
\frac{\dot{\theta}^2}{4E_c}+\frac12
\int_{-\beta/2}^{\beta/2} d\tau \int_{-\beta/2}^{\beta/2}
d\tau'\alpha(\tau-\tau')\cos[\omega_k(\tau
-\tau')][\theta(\tau)-\theta(\tau')]^2.
\]
In terms of the Fourier transform $\theta_l=
\theta'_l+i\theta''_l$,
the action takes the form
\[
\delta^2S^{(k)}= {1\over\beta}
\sum_{l=1}^{\infty}
\lambda_l^{(k)} ({\theta'_l}^2+{\theta''_l}^2) ,
\]
where for $l \ll \beta D$
\[
\lambda_l^{(k)}= \frac{\omega_l^2}{2 E_c}+\Theta(l-|k|)
\frac{\alpha_t \, \omega_{l-|k|}}{2 \pi}  .
\]
For very large $l$ the eigenvalues approach
$\omega_l^2/2 E_c$ independent of $k$.  Upon
normalization of the path
integral the large $l$ contributions cancel.  It is now
straightforward to show that
\begin{equation}
Z=Z_0 \left[ 1+2 \sum_{k=1}^{\infty} C_k
\cos (2\pi k n_{\rm ex}) \right] ,
\label{partfun}
\end{equation}
where $Z_0$ is the contribution of paths with winding number 0
which is independent of $n_{\rm ex}$ and
\begin{equation}
C_k=\frac{\Gamma (1+k_+) \Gamma (1+k_-)}{\Gamma^2 (1+k)
\Gamma (1+u)} e^{-S_{cl}^{(k)}} , \label{cock}
\end{equation}
where $u=\alpha_t\beta E_c/2\pi^2$ and $k_{\pm}=k+\frac{u}{2}\pm
\frac{1}{2} \sqrt{4uk+u^2}$.
At high temperature this result is of the form of the classical
 result
\begin{equation}
Z=\sum_{n=-\infty}^{\infty} \exp[-\beta E_c(n-n_{\rm ex})^2]
\label{zn}
\end{equation}
yet with a renormalized charging energy
\[
E_c^*=E_c \{ 1-\alpha_t \beta E_c/2 \pi^2
+{\cal O}[(\beta E_c)^2] \} .
\]

The semiclassical evaluation of the partition function
is well--behaved for all $\alpha_t$ if $ \beta E_c \ll \pi^2$.
However,
at low temperatures the ``time  scale'' $\beta$ is  large and
the eigenvalues $\lambda_l^{(k)}$ for winding number $k$
are nearly zero for small $l \leq |k|$.  Thus, a simple
semiclassical
approximation with Gaussian fluctuations around  the
classical paths becomes obsolete.   On the other hand,
a vanishing eigenvalue indicates  free motion of the minimal
action trajectory in the direction of the corresponding
eigenfunction.
Hence, we expect that at low temperatures  our system has a
 family of
trajectories  of almost the same  action.

Let us focus on the case on $k=1$ where
$\varphi(\beta/2)=\varphi(-\beta/2)+2\pi$ and
consider trajectories which gain a phase change of $2\pi$
within the imaginary time interval $s\leq\beta$.  For given $s$
there is one trajectory with smallest action.
This action is found to be almost independent
of $s$ as long as $s\gg 1/E_c$. For
$s=\beta$  one recovers  the classical path
$\varphi_{\rm cl}^{(1)}$.
Hence, there is a family of trajectories that have
nearly the same action.
In contrast to the familiar instanton trajectories
determining the low
temperature behavior of multi-well systems,
the trajectories considered
here do not have a well-defined width,
rather their action becomes smaller
as the width  grows.   In view of their
preference of a sluggish
phase change, we shall refer to these
 trajectories as ``sluggons''.

In general, the sluggon trajectories
 cannot be found analytically.
Here we consider the case
$\alpha_t \gg 1$, $\beta E_c \gg 1$ which
allows for an analytic
treatment.  In this limit multi-sluggon trajectories
with many  phase changes of $2\pi$ dominate the path integral.
The imaginary time interval $\beta$ may then
 be decomposed into segments
of length $s$ each containing a single phase change of $2\pi$.
Hence, a typical sluggon is constant outside a time
interval $s\ll\beta$
and moves within the interval $s$ along the path of least action.
Now, in the limit considered,
the kinetic part of  the action  may be treated perturbatively,
and one finds from the equation of motion
(\ref{clpath}) that \cite{kor}
\begin{equation}
\tilde{\varphi}(\tau)=2 \arctan(\Omega \tau)+\pi+
{\cal O}(1/\alpha_t, 1/{\Omega s}),
 \label{slug}
\end{equation}
is almost a minimal action path for arbitrary $\Omega$ within
the interval $1/s \ll \Omega \ll E_c$.    The sluggon action  reads
\[
S[\tilde{\varphi}(\tau)]=\frac{\alpha_t}{2}
+\frac{\pi \Omega}{2 E_c}+{\cal O}(1/\alpha_t, \alpha_t/\Omega s) ,
\]
which is very weakly dependent on
$\Omega$ as long as $\Omega \ll E_c$.

The eigenvalue problem for fluctuations $\theta(\tau)$
about the  sluggon trajectory $\tilde{\varphi}(\tau)$ reads
\[
\frac{\ddot{\theta}_n(\tau)}{2E_c}-2\int_{-s/2}^{s/2}
d\tau' \alpha(\tau-\tau') \cos [\tilde{\varphi}
(\tau)-\tilde{\varphi}(\tau')]
[\theta_n(\tau)-\theta_n(\tau')]+\Lambda_n \theta_n(\tau)=0 ,
\]
with the boundary condition  $\theta_n(-s/2)=\theta_n(s/2)$.
The eigenvalues $\Lambda_n$  $(n=1,2,3, \ldots) $ are discrete
and doubly degenerate.  In the limit  $\alpha_t \gg 1$,
$s\gg 1/\Omega \gg 1/E_c$,  the low-frequency
eigenvalues may be determined
explicitly.   We find
\[
\Lambda_n= \frac{\nu^2_n}{2E_c}+\frac{\alpha_t \nu_{n-1}}{2\pi} ,
\]
where $\nu_n=2\pi n/s$. For $n=1$, the eigenvalue is nearly zero,
and there are two linearly independent zero-modes
that may be chosen as
\[
\theta_s=\sqrt{\frac{2\Omega}{\pi}}
\frac{1}{1+(\Omega \tau)^2},
\]
and
\[
\theta_a=\sqrt{\frac{2\Omega}{\pi}}
\frac{\Omega \tau}{1+(\Omega \tau)^2}.
\]
These modes are associated with a variation of the sluggon width
and the sluggon center, respectively.
In the time interval $-s/2\leq \tau \leq s/2$, the eigenvalues
of the fluctuation modes about the trivial trajectory
$\tilde{\varphi}(\tau)=0$ are simply
\[
\Lambda^0_n=\frac{\nu^2_n}{2E_c}+ \frac{\alpha_t \nu_n}{2\pi}.
\]
The van Vleck determinant for  fluctuations about the  sluggon
with the zero-modes omitted and normalized by the fluctuation
determinant about $\tilde{\varphi}(\tau)=0$ may be
evaluated according to
\[
K=\Lambda_1^{0} \prod_{n=2}^{\infty}
\frac{\Lambda_n^{0}}{\Lambda_n} ,
\]
which gives for $ \alpha_t sE_c  \gg 1$
\[
K=\frac{\alpha_t^2}{2 \pi^2}   E_c . \]

The contribution to the partition function of paths with
winding number $k$ is the sum over all  trajectories
containing $m+k$ sluggons and $m$ anti-sluggons.
Neglecting inter-sluggon interactions, one readily finds
\begin{equation}
Z_k= \sum_{m=0}^{\infty} \frac{(2m+k)!}{(m+k)! m!} \prod_{p=1}^{2m+k}
\int_{0}^{\tau_{2p+1}}d\tau_{2p} \int_{0}^{\tau_{2p}-\tau_c}
d\tau_{2p-1} \frac{2 K e^{-\alpha_t/2}}{\tau_{2p}-\tau_{2p-1}} .
\label{pzk}
\end{equation}
The integrations over the amplitudes of the two zero-modes  of each
sluggon have been replaced by integrations over the two
``collective coordinates'' $\tau_{2p-1}$ and $\tau_{2p}$ of the sluggon,
where $(\tau_{2p-1}+\tau_{2p})/2$ is the sluggon center and
$\tau_{2p}-\tau_{2p-1}=\pi/\Omega_p$ the sluggon width.
The factor $2/(\tau_{2p}-\tau_{2p-1})$ is the Jacobian resulting from
this transformation and $\tau_c$  is the smallest sluggon  size
which is of  order $1/E_c$.

To estimate the effect of inter-sluggon interactions,
we first note that the two sluggon contribution to $Z_2$ reads
\[
Z_2^{(2)}=2 (\beta K)^2 e^{-\alpha_t} \ln^2 (\beta/2\tau_c) .
\]
This result will be modified by  inter-sluggon interactions.
For two sluggons with centers at $\tau_1$ and $\tau_2$, and
widths $\pi/\Omega_1$ and $\pi/\Omega_2$, respectively,
the asymptotic form of the interaction for large separation
is $-\alpha_t/\Omega_1 \Omega_2 (\tau_1-\tau_2)^2$.
When this is taken into account, we obtain a correction
to $Z_2^{(2)}$ proportional to $\beta^2$, hence smaller
than $Z_2^{(2)}$ by a factor of order $\ln ^{-2} (\beta/\tau_c)$.
On the other hand, for small
 separation the action of two sluggons including
interaction is bounded by the action of a $4\pi$-sluggon, which
is of order $\beta \ln (\beta/\tau_c)$.
Hence, both for small and large separation one finds a negligible
correction if $\beta$ is sufficiently large.  Likewise,
$n$ non-interacting sluggons give a leading order contribution to
$Z$ proportional to $(\beta \ln \beta)^n$ while interactions give
only corrections of smaller order for large $\beta$.  This indicates
 that the approximations made in
deriving (\ref{pzk}) are  indeed reasonable  at low temperature.

Now, from (\ref{pzk}) the  contribution for winding number $k$ reads
\begin{equation}
Z_k=\sum_{j=k}^{\infty} \frac{[2\beta K e^{-\alpha_t/2}
\ln (\beta /j \tau_c)]^j}{(\frac{j+k}{2})! (\frac{j-k}{2})!},
\label{fzk}
\end{equation}
and the partition function may be written in the form (\ref{partfun})
with $C_k=Z_k/Z_0$.  This result may be evaluated further for very
low temperatures. Let us rewrite (\ref{fzk}) as
\begin{equation}
Z_k=\sum_{j=k}^{\infty} e^{f_k(j)} , \label{zk}
\end{equation}
where
\[
f_k(x)=x \ln \left[ 2\beta K \exp(-\alpha_t/2) \ln (\beta/\tau_c x)
\right]-\ln \left[ \Gamma(x /2+k/2+1) \Gamma(x/2-k/2+1) \right]  .
\]
In the limit $\alpha_t \gg 1$, $\beta E_c \gg 1$, this function has
a maximum at $x_m= 2\beta K\alpha_t \exp (-\alpha_t/2)
[1-6\ln (\alpha_t)/\alpha_t+ {\cal O}(1/\alpha_t)]$
provided $k \ll x_m$, and
the  distribution  $\exp [f_j(x)]$ becomes Gaussian  with a width of
order  $(2 x_m)^{1/2}$.   Thus the relevant
contributions to (\ref{zk})
result from  sluggon configurations with  sluggon numbers within
$x_m-(2 x_m)^{1/2}$ and $ x_m+(2 x_m)^{1/2}$.
The mean distance
between sluggons is $\beta/x_m$, which is of order
$\alpha_t^{-3} \exp (\alpha_t/2)/E_c$.  This  is much larger than
$\tau_c$ for large $\alpha_t$, but much smaller than $\beta$ at
sufficiently  low temperatures.  Hence, the conditions under which
sluggon trajectories of the form (\ref{slug}) dominate are satisfied.

The sum  in (\ref{zk})  may  now be evaluated by means of the
Euler-Maclaurin formula \cite{abr70} yielding
\begin{equation}
C_k=\exp \left( -\pi^2 k^2/\beta E_c^*
+{\cal O}[k^4/(\beta E_c^*)^3] \right)  ,
\label{ck}
\end{equation}
where
\begin{equation}
E_c^*/E_c=2  \alpha_t^3e^{-\alpha_t/2}
[1+{\cal O}(\ln (\alpha_t)/\alpha_t)] .  \label{effec}
\end{equation}
For $n_{\rm ex}$ not near $1/2$ modulus 1,
terms with large winding number  are
irrelevant and the approximation (\ref{ck}) suffices to determine the
partition function, which is found to be of the
classical form (\ref{zn})
yet with the charging energy $E_c^*$ defined in (\ref{effec}).
Hence the quantum corrections may be absorbed in an effective
charging energy.  The exponential dependence of
$E_c^*$ on $\alpha_t$
is in accordance  with the renormalization
group analysis\cite{fal95}.
The approach presented here determines the
preexponential factor explicitly
for large tunneling conductance.
The result (\ref{effec}) differs from
earlier predictions\cite{pan91} for
the effective charging energy of
single junctions.

In summary we have exploited semiclassical methods to evaluate the
partition function of the SEB. We have obtained the leading order
quantum corrections to the effective charging energy for arbitrary
tunneling conductance  $\alpha_t$ at high temperatures.
At low temperatures the
renormalized charging energy was determined in the limit of a large
tunnel conductance. The limit of small tunneling
conductance has been treated previously\cite{gra94}. Combining
these results one sees that there remains a lacuna at intermediate
$\alpha_t$ where further work is needed. The result (\ref{effec})
can be extended to smaller values of $\alpha_t$ by avoiding the
Euler-Maclaurin expansion and evaluating (\ref{fzk}) numerically,
however, this is not sufficient to close the gap
between the available small and large conductance
results.

The authors would like to thank M.~H.~Devoret, D.~Esteve, G.~Falci,
G.~Sch{\"o}n, and W.~Zwerger for fruitful discussions.
Financial support was provided by the
Deutsche Forschungs\-ge\-mein\-schaft (Bonn) and
the European Community under Contract No. SC1$^*$CT91-0631.


\begin{references}
\bibitem{gra92} Single Charge Tunneling,  NATO ASI Series B294,
ed. by H. Grabert and  M. H. Devoret (Plenum, New York, 1992).
\bibitem{ing92} G.-L. Ingold and Yu. V. Nazarev, in Ref. 1.
\bibitem{mat95} K. Flensberg, Phys.\ Rev.\ B {\bf 48}, 11156 (1993);
K. A. Matveev,  Phys.\ Rev.\ B {\bf 51}, 1743 (1995).
\bibitem{ben83} E. Ben-Jocab, E. Mottola, and G. Sch{\"o}n,
Phys. Rev. Lett. {\bf 51}, 2064 (1983);
A. D. Zaikin and G. Sch{\"o}n, Phys. Rep. {\bf 198}, 237 (1990).
\bibitem{gui86} F. Guinea and  G. Sch{\"o}n,
Europhys. Lett. {\bf 1}, 585(1986).
\bibitem{pan91} S. V. Panyukov, and A. D. Zaikin,
Phys. Rev. Lett. {\bf 67}, 3168 (1991).
\bibitem{zw91}   W. Zwerger and M. Scharpf,
Z. Phys. B{\bf 85} (1991).
\bibitem{fal95} G. Falci, G. Sch{\"o}n, and G. T.\ Zimanyi,
Phys. Rev. Lett. {\bf 74}, 3257 (1995).
\bibitem{dev90} M. H. Devoret, D. Esteve, H. Grabert,
G.-L. Ingold, H. Pothier, and C. Urbina,
Phys.\ Rev.\ Lett.\ {\bf 64}, 1824 (1990);
S. M. Girvin, L. I. Glazman, M. Jonson, D. R. Penn,
and M. D. Stiles   {\sl ibid.} {\bf 64}, 3183 (1990)
\bibitem{av92} D. V. Averin and Yu. V. Nazarev, in Ref. 1.
\bibitem{mat91} K. A. Matveev, Zh.\ Eksp.\ Teor.\ Fiz.
{\bf 99}, 1598 (1991) [Sov.\ Phys. JETP {\bf 72}, 892 (1991).
\bibitem{gog94} D.~S. Golubev and A.~D. Zaikin,
Phys.\ Rev.\ B {\bf 50}, 8736 (1994)
\bibitem{gra94} H. Grabert,   Phys. Rev. B{\bf 50},  17364 (1994).
\bibitem{sch94}
H.~Schoeller and G.~Sch{\"o}n,
Phys.\ Rev.\ B {\bf 50}, 18436 (1994)
\bibitem{laf93} P. Lafarge, H. Pothier, E. R. Williams, D. Esteve,
C. Urbina and M. H. Devoret,
Z. Phys.\ B {\bf 85}, 327 (1991).
\bibitem{kor} S. E. Korshunov,  Pis'ma Zh. Eksp. Teor. Fiz.
{\bf 45}, 342 (1987) [JETP Lett. {\bf 45}, 434(1987)].
\bibitem{abr70} Handbook of Mathematical Functions,
ed. by M.\ Abramowitz and I. A.\ Stegun (Dover, New York, 1970).
\end{references}
\end{document}